\documentclass[a4paper,12pt]{article}
\usepackage{graphics}
\global\arraycolsep=2pt 
\newlength{\orbarwd}\newlength{\orbarht}\newsavebox{\orbararg}%
\newcommand{\orbar}[1]{\savebox{\orbararg}{\ensuremath{#1}}%
  \settowidth{\orbarwd}{\usebox{\orbararg}}%
  \settoheight{\orbarht}{\usebox{\orbararg}}%
  \raisebox{1.1\orbarht}[0pt]{\makebox[0pt][l]{%
    \resizebox{1.1\orbarwd}{0.5ex}{\boldmath\ensuremath{(-)}}}}%
  \usebox{\orbararg}}%
\newenvironment{darray}[2][]{
  \newcommand{\nc}{&\displaystyle}
  \newcommand{\nl}{\\\displaystyle}
  \begin{array}[#1]{#2}\displaystyle}{\end{array}}%
\newenvironment{deqnarray}
  {\begin{equation}\begin{darray}{rcl}}{\end{darray}\end{equation}}%
\begin{document}
\makeatletter
\def\fmslash{\@ifnextchar[{\fmsl@sh}{\fmsl@sh[0mu]}}
\def\fmsl@sh[#1]#2{%
  \mathchoice
    {\@fmsl@sh\displaystyle{#1}{#2}}%
    {\@fmsl@sh\textstyle{#1}{#2}}%
    {\@fmsl@sh\scriptstyle{#1}{#2}}%
    {\@fmsl@sh\scriptscriptstyle{#1}{#2}}}
\def\@fmsl@sh#1#2#3{\m@th\ooalign{$\hfil#1\mkern#2/\hfil$\crcr$#1#3$}}
\makeatother
%
\thispagestyle{empty}
\begin{titlepage}

\begin{flushright}
Phys.\ Rev.\ {\bf D61}, 114004 \\
hep-ph/9907378 \\
TTP99--31 \\
1 May 2000
\end{flushright}

\vspace{0.3cm}
\boldmath
\begin{center}
\Large\bf QCD-based description of \\ 
          one-particle inclusive $B$ decays           
\end{center}
\unboldmath
\vspace{0.8cm}

\begin{center}
{\large Xavier Calmet}, {\large Thomas Mannel}, and {\large Ingo 
Schwarze} \\
{\sl Institut f\"{u}r Theoretische Teilchenphysik,
     Universit\"at Karlsruhe,\\ D--76128 Karlsruhe, Germany}

\vspace{\fill}
Published in Physical Review {\bf D61}, 114004 \\
(Received 16 July 1999; published 1 May 2000)
\end{center}

\vspace{\fill}

\begin{abstract}
\noindent
We discuss one-particle inclusive $B$ decays in the limit of heavy $b$
and $c$ quarks.  Using the large-$N_C$ limit we factorize the
nonleptonic matrix elements, and we employ a short distance
expansion.  Modeling the remaining nonperturbative matrix elements we
obtain predictions for various decay channels and compare them with
existing data.

\medskip\noindent PACS numbers: 13.25.Hw, 12.39.Hg
\end{abstract}
\end{titlepage}

\setcounter{page}{2}
\thispagestyle{empty}
\mbox{}
\clearpage

\setcounter{page}{1}
\section{Introduction}
Over the last ten years, methods have been developed to describe the
decays of heavy hadrons within a framework based on the heavy mass
expansion of QCD.  Using this approach, model dependences have been
drastically reduced.

The heavy mass expansion has been formulated for exclusive as well as
for inclusive decays.  While in the former case the heavy mass
expansion is the well-known formalism of heavy quark effective theory
(HQET) \cite{HQET}, the latter case additionally requires a short
distance expansion (SDE) \cite{OPE} which is very similar to the
operator product expansion (OPE) used for deep inelastic scattering.
Both approaches are QCD based methods and have a good theoretical
foundation.

However, nothing comparable exists for for a theoretical description
of one-particle inclusive decays such as $B \to \orbar{D} X$ or $B \to
\orbar{K} X$.  Unlike the fully inclusive case, the expression for the
rate always involves a projection on a specific particle, thereby
spoiling a straightforward short distance expansion. The
semileptonic case has been discussed in \cite{BM}, and the
present paper is devoted to a study of the nonleptonic decays.

We shall concentrate on decays of the form $B \to \overline D X$ and
$B \to DX$.  These decays have already been considered long ago in the
context of models \cite{Wirbel89}.  Here we aim at a QCD description
and exploit the heavy mass limit for both the $b$ and the $c$ quark.
Furthermore we shall use the large-$N_C$ limit to factorize the
hadronic matrix element.  In this way we can identify the parts which
allow a short distance expansion.  The remaining contributions have to
be parametrized and we shall discuss simple forms of this
parametrization.

We first discuss the right charm contribution arising from quark level
$b \to c X$ decays and the wrong charm contribution arising from $b
\to \overline c X$ separately.  After recalling some facts about the
phenomenology of $\orbar{D}$ mesons in Sec.~\ref{secD}, we predict
rates and spectra for various decay channels in Sec.~\ref{results}.
Finally we compare the results with existing data.

\section{The effective Hamiltonian\\
         and the right charm contribution}\label{secRC}

The relevant effective Hamiltonian for the decays $B \to \orbar{D} X$
is given by
\begin{equation}
  H_{eff} = H_{eff}^{(sl)} + H_{eff}^{(nl)},
\end{equation}
where the semileptonic and nonleptonic pieces are
\begin{equation}
  H_{eff}^{(sl)} = \frac{G_F}{\sqrt{2}} \, V_{cb} \,
    (\overline b c)_{V-A} \, (\overline\ell \nu)_{V-A} + {\rm H.c.}, \\
\end{equation}
\begin{equation}
  H_{eff}^{(nl)} \! = \! \frac{G_F}{\sqrt{2}} \!
    \sum_{q=d,s} \!\!\! \left( \! 
    V^{ }_{\!cb}\! V^*_{\!uq}\! \sum_{k=1}^{2}\!\! C_k(\mu) O_k^q\!+\!
    V^{ }_{\!cb}\! V^*_{\!cq}\! \sum_{k=1}^{2}\!\! C_k(\mu) P_k^q\!+\!
    V^{ }_{\!tb}\! V^*_{\!tq}\! \sum_{k=3}^{6}\!\! C_k(\mu) O_k^q\!
    \right) \! + {\rm H.c.}
\end{equation}
$O_{1,2}$ and $P_{1,2}$ are the current-current operators, and $O_{3
\ldots 6}$ are the QCD penguin operators.  For a full list of
operators, see, e.g., the review \cite{Buchalla96}.  We shall
only consider the Cabibbo-favored decays and neglect penguin
contributions. Hence the operators we need are
\begin{deqnarray}
  O^d_1 = O_1 & = \nc (\overline b c)_{V-A} (\overline u d)_{V-A} \nl
  O^d_2 = O_2 & = \nc (\overline b T^{a} c)_{V-A}
                      (\overline u T^{a} d)_{V-A} \nl
  P^s_1 = P_1 & = \nc (\overline b c)_{V-A} (\overline c s)_{V-A} \nl
  P^s_2 = P_2 & = \nc (\overline b T^{a} c)_{V-A}
                      (\overline c T^{a} s)_{V-A}.
\end{deqnarray}%
$C_1$ and $C_2$ are the Wilson coefficients encoding the short
distance physics and $T^a$ are the generators of color-SU(3).  The
operators $O_1$ and $O_2$ as well as the semileptonic Hamiltonian
$H_{eff}^{(sl)}$ contribute to right charm transitions only, while
$P_1$ and $P_2$ contribute to both right and wrong charm processes.

It is well known that in the large-$N_C$ limit the matrix elements of
the four-fermion operators factorize into products of two current
matrix elements \cite{Hooft74}.
Factorization has also been investigated on a phenomenological basis
and found to work well for exclusive
nonleptonic decays of $B$ mesons \cite{BSW}.  The contributions of
$O_2$ and $P_2$ vanish in the factorization limit since the currents
are color octets.  Furthermore, the coefficient $C_1$ differs from
unity only through radiative corrections which are very small and will
be neglected.

The semileptonic case has already been studied in \cite{BM}, so we
focus on the nonleptonic modes.  We consider the matrix element
\begin{equation}
  G(M^2) = \sum_X \left| \langle \overline D^{(*)}(p_{\overline D}) X 
           | H_{eff} | B(p_B) \rangle \right|^2
           (2 \pi)^4 \delta^4 (p_B - p_{\overline D} - p_X)
\end{equation}
where the states $| X \rangle$ form a complete set of momentum
eigenstates with momentum $p_X$ and
$H_{eff}$ is the relevant part of the weak Hamiltonian. The function
$G$ depends on the invariant mass
\begin{equation}
  M^2 = (p_B - p_{\overline D})^2
\end{equation}
of the state $|X\rangle$.  It is related to the decay rate under
consideration by
\begin{equation}
  d\Gamma(B \to \overline D^{(*)} X) =
    \frac{1}{2 m_B} \, d\Phi_{\overline D} \; G(M^2),
\end{equation}
where $d\Phi_{\overline D}$ is the phase space element of the final
state $\overline D^{(*)}$ meson.
     
\begin{figure}
\begin{center}
\includegraphics{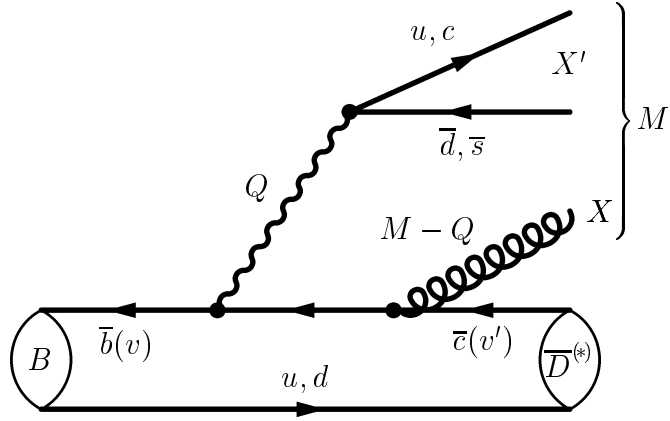}
\caption{Momenta and velocities used for the description of the right
charm contribution.}
\label{figRC}
\end{center}
\end{figure}

Due to the different final states $c \overline u d$ and $c
\overline c s$, there are no interference terms between $O_1$ and
$P_1$.  The contribution of the channel $b \to c \overline u d$ is
\begin{eqnarray}
  G_1 (M^2) & = &
    \frac{G_F^2}{2} \, |V_{cb} V_{ud}^*|^2 \, |C_1|^2 \;
      \sum_X (2 \pi)^4 \delta^4 (M - p_X) \\
    \nonumber &&
    \langle B(p_B) | (\overline c \gamma_\mu (1-\gamma_5) b) 
                     (\overline d \gamma^\mu (1-\gamma_5) u)
                   | \overline D^{(*)}(p_{\overline D}) X \rangle \\
    \nonumber &&
    \langle \overline D^{(*)}(p_{\overline D}) X | 
           (\overline u \gamma^\nu (1-\gamma_5) d)
           (\overline b \gamma_\nu (1-\gamma_5) c) | B(p_B) \rangle.
\end{eqnarray}
Using the factorization of the large-$N_C$ limit we have (see
Fig.~\ref{figRC})
\begin{eqnarray}
  G_1(M^2) & = &
    \frac{G_F^2}{2} \, |V_{cb} V_{ud}^*|^2 \, |C_1|^2 \;
      \sum_X \sum_{X'} (2 \pi)^4 \delta^4 (M - p_X - p_{X'}) \\
    \nonumber &&
    \langle B(p_B) | (\overline c \gamma_\mu (1-\gamma_5) b) 
                   |  \overline D^{(*)}(p_{\overline D}) X \rangle \;
      \langle 0    | (\overline d \gamma^\mu (1-\gamma_5) u)
                   | X' \rangle \\
    \nonumber &&
    \langle X' | (\overline u \gamma^\nu (1-\gamma_5) d) | 0 \rangle \;
       \langle \overline D^{(*)}(p_{\overline D}) X
          | (\overline b \gamma_\nu (1-\gamma_5) c) | B(p_B) \rangle.
\end{eqnarray}
It is convenient to define two tensors
\begin{eqnarray}
  \lefteqn{K_{\mu \nu} (p_B,M,Q) =
           \sum_X (2 \pi)^4 \delta^4 (M - Q - p_X)} \\
  \nonumber 
  && \quad \langle B(p_B) | (\overline c \gamma_\mu (1-\gamma_5) b) 
                  | \overline D^{(*)}(p_{\overline D}) X  \rangle 
           \langle  \overline D^{(*)}(p_{\overline D}) X |  
                   (\overline b \gamma_\nu (1-\gamma_5) c) 
                  | B(p_B)  \rangle 
\end{eqnarray}
and 
\begin{equation}
  P_{\mu \nu}(Q) = \sum_{X'} (2 \pi)^4 \delta^4 (Q - p_{X'})
    \langle 0  | (\overline d \gamma_\mu (1-\gamma_5) u) | X' \rangle
    \langle X' | (\overline u \gamma_\nu (1-\gamma_5) d) | 0  \rangle 
\end{equation}
in terms of which we obtain for the rate 
\begin{equation}
  G_1(M^2) = \frac{G_F^2}{2} \, |V_{cb} V_{ud}^*|^2 \, |C_1|^2 
  \int \frac{d^4 Q}{(2\pi)^4} \,
  K_{\mu \nu} (p_B,M,Q) \, P^{\mu \nu}(Q).
\end{equation}

The tensor $P_{\mu \nu}$ involves only light quarks and can be
rewritten as
\begin{equation}
  P_{\mu \nu}(Q) = \int d^4 x \, e^{-iQx}
    \langle 0 | (\overline d(x) \gamma_\mu (1-\gamma_5) u(x))
                (\overline u(0) \gamma_\nu (1-\gamma_5) d(0))
              | 0 \rangle.
\end{equation}
For sufficiently large $Q$ this quantity has a short distance
expansion in inverse powers of $Q$.
However, the momentum $Q$ is not a measurable kinematical
quantity.  In particular, it is not equal to the recoil $M$, but in
most of the phase space, it should be of the same order as $M$, namely
${\cal O}(m_b-m_c)$.  Close to the nonrecoil point, $M$ is large,
corresponding to either large $Q$ or a large momentum of the gluon
depicted in Fig.~\ref{figRC}.  While in the former case the OPE
treatment is justified, the latter case could be treated
perturbatively.  In other words, the two expansion parameters for the
total rate are $\Lambda_{QCD}/(m_b-m_c)$ and $\alpha_s(m_c)/\pi$.
However we also consider the momentum spectra of the $\orbar{D}$
mesons.  In that case the expansion parameter depends on the energy
of the $\orbar{D}$ meson since
\begin{equation}
  Q^2 \approx M^2 = ( m_b - m_c )^2 - 2 m_b ( E_c - m_c ).
\end{equation}

The leading term of the short distance expansion yields the partonic
result
\begin{equation}\label{Ppart}
  P_{\mu \nu}(Q) =
  \frac{N_C}{3 \pi} \, (Q_\mu Q_\nu - g_{\mu \nu} Q^2) \, \Theta(Q^2)
\end{equation}
where we have assumed both light quarks to be massless. 

For the quantity $K_{\mu \nu}$ it is convenient to use the heavy mass
limit for both the bottom and the charm quark. To this end we redefine
the phases of the quark fields as
\begin{equation}\label{resca}
  b(x) = b_v(x)    \, e^{-im_b vx}, \qquad
  c(x) = c_{v'}(x) \, e^{-im_c v'x},
\end{equation}
where the velocities are defined as $p_B = m_B v$ and $p_{\overline D}
= m_{\overline D} v'$. In the following we shall work in the infinite
mass limit for the $b$ and the $c$ quark, and we obtain
\begin{eqnarray}
  K_{\mu \nu}(m_b v,M,Q) & = &
    \int d^4 z \, \sum_X \exp[-i(M - Q)z] \\
    \nonumber &&
    \langle B(v)
      | (\overline c_{v'}(z) \gamma_\mu (1-\gamma_5) b_v(z))
      |  \overline D^{(*)}(v') X \rangle \\
    \nonumber &&
    \langle \overline D^{(*)}(v') X
      | (\overline b_v(0) \gamma_\nu (1-\gamma_5) c_{v'} (0))
      | B(v) \rangle,
\end{eqnarray}
where in the heavy mass limit $M$ becomes $m_b v - m_c v'$.

Using the result in Eq.~(\ref{Ppart}) for $P_{\mu \nu}$ we can express
the rate in terms of the two quantities

\begin{eqnarray}
  K_1(v,v',Q) &=& \frac{Q_\mu Q_\nu}{Q^2} K^{\mu \nu}(m_b v,M,Q), \\
  K_2(v,v',Q) &=& K_{\ \mu}^\mu (m_b v,M,Q),
\end{eqnarray}
and obtain 
\begin{eqnarray}
  G_1(M^2) & = &
    \frac{G_F^2}{2} \, |V_{cb} V_{ud}^*|^2 \, |C_1|^2 \,
    \frac{N_C}{3 \pi} \\
  \nonumber &&
  \int \frac{d^4 Q}{(2\pi)^4} \, Q^2 \; \Theta(Q^2) \,
    [K_1 (v,v',Q) - K_2 (v,v',Q)].
\end{eqnarray}

Not much can be said about the functions $K_1$ and $K_2$. They do not
have an obvious short distance expansion due to the projection on the
final state $\overline D$ meson. The only restriction we have is from
the spin symmetry of the heavy $b$ and $c$ quarks. To implement these
symmetries we use the spin projection matrices for the heavy mesons
\begin{deqnarray}
  H_B(v) & = \nc \frac{1}{2} \sqrt{m_B} (1+\fmslash{v}) \gamma_5, \nl
  H_{\overline D^{(*)}}(v') & = \nc
    \frac{1}{2} \sqrt{m_D} (1+\fmslash{v}') 
    \left\{ \begin{array}{cl}
              \gamma_5 & \mbox{(pseudoscalar meson),} \\
              \fmslash{\epsilon} & \mbox{(vector meson).}
            \end{array} \right.
\end{deqnarray}%
In the heavy mass limit, $D$ and $D^*$ become degenerate and
constitute the ground state spin symmetry doublet of the $D$ meson
system.  For phenomenological applications we shall later take the
mass splitting between the $D$ and $D^*$ mesons into account, although
formally this is a $1/m_c$ effect.

As far as the spinor indices are concerned, the tensor $K_{\mu \nu}$
is given by
\begin{equation}\label{eqContr}
  K_{\mu \nu}(p_B,M,Q) \propto 
  \overline H_B (v)  \gamma_\mu (1-\gamma_5)
    H_{\overline D^{(*)}} (v')  \otimes 
  \overline H_{\overline D^{(*)}} (v')
    \gamma_\nu (1-\gamma_5)  H_B (v),
\end{equation}
where the remaining indices are light quark indices which have to be
contracted using the most general four-index object.

Since there are many possibilities to contract the indices, the
discussion of the general case would leave us with a large number of
unknown functions, so we have to make a choice.  The matrix elements
appearing in $K_{\mu \nu}$ are identical to the semileptonic case.
Therefore we shall use the same ansatz for the nonleptonic as for the
semileptonic case \cite{BM} and write
\begin{eqnarray}\label{factrc}
  \lefteqn{K_{\mu \nu}(p_B,M,Q) =
    (2 \pi)^4 \delta^4 (M - Q) \; \eta(vv')} \\
  \nonumber
  && {\rm Tr}[\overline H_B(v) \gamma_\mu (1-\gamma_5)
       H_{\overline D^{(*)}}(v')] \;
     {\rm Tr}[\overline H_{\overline D^{(*)}}(v')
       \gamma_\nu (1-\gamma_5) H_B (v)].
\end{eqnarray}
In this way we can model the two functions $K_1$ and $K_2$ in terms of
the single nonperturbative function $\eta(vv')$, the advantage being
that spin symmetry relates the rates of $B \to \overline D X$ and $B
\to \overline D^*X$ to each other.

Finally we also have to consider the right charm contribution of the
operator $P_1$, which is
\begin{eqnarray}
  G_2(M^2) & = &
    \frac{G_F^2}{2} \, |V_{cb} V_{cs}^*|^2 \, |C_1|^2 \;
      \sum_X (2 \pi)^4 \delta^4 (M - p_X) \\
  \nonumber &&
  \langle B(p_B) | (\overline c \gamma_\mu (1-\gamma_5) b) 
                   (\overline s \gamma^\mu (1-\gamma_5) c)
                 |  \overline D^{(*)}(p_{\overline D}) X \rangle \\
  \nonumber &&
  \langle \overline D^{(*)}(p_{\overline D}) X | 
         (\overline c \gamma^\nu (1-\gamma_5) s)
         (\overline b \gamma_\nu (1-\gamma_5) c) | B(p_B) \rangle \;.
\end{eqnarray}
The calculation is exactly the same as before, but the short distance
expansion of $P_{\mu \nu}$ yields a different result since now a
massive charm quark is involved. We have
\begin{equation}
  P'_{\mu \nu}(Q) = \sum_{X'} (2 \pi)^4 \delta^4 (Q - p_{X'}) 
    \langle 0  | (\overline s \gamma_\mu (1-\gamma_5) c) | X' \rangle
    \langle X' | (\overline c \gamma_\nu (1-\gamma_5) s) | 0  \rangle 
\end{equation}
from which we have the leading order contribution
\begin{equation}\label{pheavy}
  P'_{\mu \nu}(Q) =
    \left( A(Q^2) \, Q_\mu Q_\nu - B(Q^2) \, Q^2 g_{\mu \nu} \right)
    \Theta(Q^2-m_c^2),
\end{equation}
with 
\begin{deqnarray}
  A(Q^2) & = \nc \frac{N_C}{3 \pi}
                 \left(1 - \frac{m_c^2}{Q^2} \right)^2
                 \left(1 + 2 \frac{m_c^2}{Q^2} \right), \nl
  B(Q^2) & = \nc \frac{N_C}{3 \pi}
                 \left(1 - \frac{m_c^2}{Q^2} \right)^2
                 \left(1 + \frac{1}{2} \frac{m_c^2}{Q^2} \right),
\end{deqnarray}%
and we get 
\begin{eqnarray}\label{rcfinal}
  G_2 (M^2) & = & \frac{G_F^2}{2} \, |V_{cb} V_{cs}^*|^2 \, |C_1|^2 \;
    \int \frac{d^4 Q}{(2\pi)^4} \, \Theta(Q^2-m_c^2) \; Q^2 \\
  \nonumber &&
  [A(Q^2) K_1 (v,v',Q) - B(Q^2) K_2 (v,v',Q)].
\end{eqnarray}
Thus the same two nonperturbative functions appear in $G_2$.

\section{The wrong charm contribution}
\begin{figure}
\begin{center}
\includegraphics{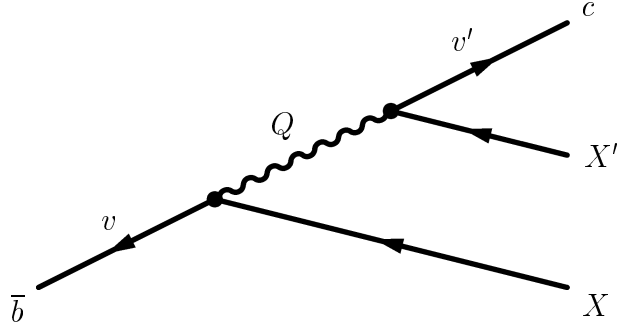}
\caption{Momenta and velocities used for the description of the wrong
charm contribution.}
\label{figWC}
\end{center}
\end{figure}

Wrong charm decays can only be mediated by the operators $P_1$ and
$P_2$. In the large-$N_C$ limit, the contribution of $P_2$ can be
neglected and we have
\begin{eqnarray}
  G_3(p_B,p_D) & = &
    \frac{G_F^2}{2} \, |V_{cb} V_{cs}^*|^2 \, |C_1|^2 \;
    \sum_X (2 \pi)^4 \delta^4 (p_B - p_D - p_X) \\
  \nonumber &&
  \langle B(p_B) | (\overline c \gamma_\mu (1-\gamma_5) b) 
                   (\overline s \gamma^\mu (1-\gamma_5) c)
                 | D^{(*)}(p_D) X \rangle \\
  \nonumber &&
  \langle D^{(*)}(p_D) X | (\overline c \gamma^\nu (1-\gamma_5) s)
    (\overline b \gamma_\nu (1-\gamma_5) c) | B(p_B) \rangle.
\end{eqnarray}
Again using the factorization of the large-$N_C$ limit we get (see
Fig.~\ref{figWC})
\begin{eqnarray}
  G_3(p_B,p_D) & = &
    \frac{G_F^2}{2} \, |V_{cb} V_{cs}^*|^2 \, |C_1|^2 \;
    \sum_X \sum_{X'} (2 \pi)^4 \delta^4 (p_B - p_D - p_X - p_{X'}) \\
  \nonumber &&
  \langle B(p_B) | (\overline c \gamma_\mu (1-\gamma_5) b)
                 | X \rangle \;
    \langle X | (\overline b \gamma_\nu (1-\gamma_5) c)
              | B(p_B) \rangle \\
  \nonumber &&
  \langle 0 | (\overline s \gamma^\mu (1-\gamma_5) c)
            | D^{(*)}(p_D) X' \rangle \;
    \langle D^{(*)}(p_D) X' | (\overline c \gamma^\nu (1-\gamma_5) s) 
                            | 0 \rangle.
\end{eqnarray}
In a similar way as before it is convenient to define two tensors
\begin{eqnarray}
  K'_{\mu \nu} (p_B,Q) &=& \sum_X (2 \pi)^4 \delta^4 (p_B - p_X - Q) \\
  \nonumber &&
    \langle B(p_B) | (\overline c \gamma_\mu (1-\gamma_5) b)| X \rangle
    \langle X | (\overline b \gamma_\nu (1-\gamma_5) c)| B(p_B) \rangle
\end{eqnarray}
and 
\begin{eqnarray}
  R_{\mu \nu}(p_D,Q) & = &
    \sum_{X'} (2 \pi)^4 \delta^4 (Q - p_D - p_{X'}) \\
  \nonumber && 
  \langle 0 |  (\overline s \gamma_\mu (1-\gamma_5) c) 
            | D^{(*)}(p_D) X' \rangle 
  \langle D^{(*)}(p_D) X' | (\overline c \gamma_\nu (1-\gamma_5) s) 
            | 0 \rangle,
\end{eqnarray}
in terms of which the rate becomes 
\begin{equation}
  G_3(p_B,p_D) = \frac{G_F^2}{2} \, |V_{cb} V_{cs}^*|^2 \, |C_1|^2 \;
    \int \frac{d^4 Q}{(2\pi)^4} \,
    K'_{\mu \nu}(p_B,Q) \, R^{\mu \nu}(p_D,Q).
\end{equation}

The quantity $K'_{\mu \nu}$ is fully inclusive and one may perform a
short distance expansion.  In the heavy mass limit for the $b$ quark
it is convenient to rescale the $b$ quark field as in
Eq.~(\ref{resca}), and we obtain
\begin{eqnarray}
  K'_{\mu \nu}(p_B,Q) & = & \int d^4 y \, \exp[-i(m_b v - Q)y] \\
  \nonumber &&
  \langle B(v) | (\overline c(y) \gamma_\mu (1-\gamma_5) b_v(y)) 
                 (\overline b_v(0)   \gamma_\nu (1-\gamma_5) c(0)) 
               | B(v) \rangle.
\end{eqnarray} 
In the region where the momentum $m_b v - Q$ is large, we can perform
a short distance expansion.  As before, $m_b v - Q$ is not an
observable, but it is of the order of $m_b-m_c$ in most of the phase
space.  The leading term is the dimension three operator $\overline
b_v b_v$, the matrix elements of which are normalized due to heavy
quark symmetry. Thus we obtain
\begin{eqnarray}
  K'_{\mu \nu}(m_b v,Q) & = & 2 \pi
    \delta \left( (m_b v - Q)^2 - m_c^2 \right) \\
  \nonumber &&
  {\rm Tr}[\overline H_B(v) \gamma_\mu (1-\gamma_5)
           (m_b \fmslash{v} - \fmslash{Q})
           \gamma_\nu (1-\gamma_5) H_B(v)].
\end{eqnarray}
The other factor $R_{\mu \nu}$ involves a projection on a 
$D^{(*)}$ meson in the intermediate state and has to be parametrized. 
Heavy quark spin symmetry for the $c$ quark implies that 
$R_{\mu \nu}$ is of the form 
\begin{equation}
  R_{\mu \nu}(m_c v',Q) \propto 
    \overline H_{D^{(*)}}(v') \, \gamma_\mu (1-\gamma_5) \otimes 
    \gamma_\nu (1-\gamma_5) \, H_{D^{(*)}}(v'),
\end{equation}
where the remaining light quark indices have to be contracted 
using the most general four-index object.

Again there is quite a large number of possibilities to contract the
indices, making it useless to discuss the general case. Hence we shall
only give two physically motivated ways for modeling this quantity.

The first model ansatz corresponds to factorization:
\begin{eqnarray}\label{factwc}
  R_{\mu \nu} (m_c v',Q) & = & 2 \, F(v'Q,Q^2) \\
  \nonumber &&
  {\rm Tr} [\overline H_{D^{(*)}} (v') \, \gamma_\mu (1-\gamma_5)] \;
  {\rm Tr} [ \gamma_\nu (1-\gamma_5) \, H_{D^{(*)}} (v') ],
\end{eqnarray}
the second one is inspired by the parton model and is defined through
\begin{eqnarray}\label{partwc}
  R_{\mu \nu}(m_c v',Q) & = & 2 \, \tilde{F}(v'Q,Q^2) \\
  \nonumber &&
  {\rm Tr} [ \overline H_{D^{(*)}} (v') \gamma_\mu (1-\gamma_5) 
             (\fmslash{Q} - m_c \fmslash{v}')
             \gamma_\nu (1-\gamma_5) H_{D^{(*)}} (v') ].
\end{eqnarray}
We shall discuss the functions $F$ and $\tilde{F}$ using data.

\section{\boldmath Effects of $D^* \to D$ decays}\label{secD}
The right charm decays of $b = +1$ mesons are the ones into $\overline
D^0$ or $D^-$
\begin{equation}
  \begin{array}{cccc}
    B^+ & \to & \overline D^{(*)0} & X, \\
    B^+ & \to & D^{(*)-} & X,
  \end{array}
  \qquad
  \begin{array}{cccc}
    B^0 & \to & \overline D^{(*)0} & X, \\
    B^0 & \to & D^{(*)-} & X,
  \end{array}
\end{equation} 
while the wrong charm decays are
\begin{equation}
  \begin{array}{cccc}
    B^+ & \to & D^{(*)0} & X, \\
    B^+ & \to & D^{(*)+} & X,
  \end{array}
  \qquad
  \begin{array}{cccc}
    B^0 & \to & D^{(*)0} & X, \\
    B^0 & \to & D^{(*)+} & X.  
  \end{array}
\end{equation} 
For the semileptonic case the charge of the the lepton tags the $b$
flavor of the decaying $B$ meson.  Wrong charm semileptonic decays
are suppressed by the large charm mass and will be ignored in our
discussion.

Off the heavy mass limit the degeneracy between $\orbar{D}$ and
$\orbar{D}^*$ mesons is removed.  In fact the mass difference is large
enough to allow strong decays into pions.  Thus the rate for decays
into $\orbar{D}$ mesons is the sum of a direct contribution and the
contribution arising from the decay chain $B \to \orbar{D}^* X' \to
\orbar{D} X$.  In the narrow width approximation, the latter is
obtained by weighting the rate for $B \to \orbar{D}^* X$ by the
branching ratios $D^* \to D Y$ where Y is either a pion or a photon.

While $D^{* \pm}$ decays are governed by the isospin Clebsch-Gordan
coefficients receiving only tiny corrections from phase space effects
and from the radiative process,
\begin{equation}
  {\rm Br} (D^{*+} \to D^+ Y) = 0.32, \qquad 
  {\rm Br} (D^{*+} \to D^0 Y) = 0.68
\end{equation}
in $D^{*0}$ decays isospin invariance is maximally broken by
phase space effects \cite{Goldh76} such that
\begin{equation}
  {\rm Br} (D^{*0} \to D^0 Y) = 1.
\end{equation}
Consequently the total branching ratios to pseudoscalar $\overline
D$ mesons are
\begin{eqnarray}\label{feeddown}
  {\rm Br}(B \!\to\! D^- X) & = &
    {\rm Br}(B \!\to\! D^-_{\rm dir} X) +
    0.32 \, {\rm Br}(B \!\to\! D^{*-} X) \\ \nonumber
  {\rm Br}(B \!\to\! \overline D^0 X) & = &
    {\rm Br}(B \!\to\! \overline D^0_{\rm dir} X) +
    {\rm Br}(B \!\to\! \overline D^{*0} X) +
    0.68 \, {\rm Br}(B \!\to\! D^{*-} X)
\end{eqnarray}
where $\overline D_{\rm dir}$ refers to the direct contribution, i.e.,
the contribution where no $\overline D^*$ meson appears in an
intermediate state.

We shall later consider spin and charge counting in one-particle
inclusive $B$ decays and define the following ratios:
\begin{eqnarray}
  \label{spincount}
    r_S & = & \frac{{\rm Br}(B \to \orbar{D}^* X)}
                   {{\rm Br}(B \to \orbar{D}_{\rm dir} X)
                                   \rule{0pt}{2.3ex} }, \\
  \label{chargecount}
    r_Q & = & \frac{{\rm Br}(B \to \orbar{D}^0 X)}
                   {{\rm Br}(B \to D^\pm X)}.
\end{eqnarray}
We assume that $\orbar{D}^*$ decay is the only relevant isospin
violating effect.  Under this assumption the charge counting ratio is
governed by the spin counting ratio:
\begin{equation}\label{chargespin}
    r_Q = \frac{1 + 1.68 r_S}{1 + 0.32 r_S}.
\end{equation}
Assuming equal rates for the pseudoscalar mesons and for each
polarization state of the vector mesons, the naive expectations
for these two ratios are $r_S = 3$ and $r_Q = 3$.

Apart from the ground states $\orbar{D}$ and $\orbar{D}^*$, exited
$\orbar{D}$ mesons may be produced in $B$ meson decays.  Even in the
heavy mass limit a finite splitting to the ground state mesons
remains, which means that these states will decay into $\orbar{D}$ or
$\orbar{D}^*$.  In order to analyze decays into higher resonances, one
would need to study the state $|X\rangle$, which we take fully
inclusive.  Thus we may rely on parton hadron duality to obtain a
correct description of the $\orbar{D}$ and $\orbar{D}^*$ production up
to leading order in the heavy mass expansion.

Charmed mesons can also be produced via charmonium intermediate
states.  These states show up in the invariant mass distribution
$d \Gamma / d m_{D \overline D}$ as more or less pronounced resonances.
The distribution $d \Gamma / d m_{c \overline c}$ is certainly
different, but by global parton hadron duality and the assumption
that all charm quarks hadronize into $\orbar{D}$ mesons, their
integrals are equal, so the charmonium contributions are included in
our calculation.

\section{Results}\label{results}

Even after restricting the number of possible form factors by the
ans\"atze in Eqs.~(\ref{factrc}), (\ref{factwc}) and (\ref{partwc}),
we are still left with unknown nonperturbative functions.  As far as
the right charm contributions are concerned we shall first consider
the decays into $\overline D^*$ mesons.  According to
Eq.~(\ref{factrc}), these are described by a single function
$\eta(vv')$.  Spin symmetry relates these decays to the ones into
$\overline D$ mesons, however, this is only true in the heavy quark
limit, where again $\eta(vv')$ is the nonperturbative input.  In
reality one has to take into account that $\overline D^*$ mesons decay
into $\overline D$'s, and thus one has to add this contribution using
Eqs.~(\ref{feeddown}).

Still all right charm decays are given in terms of the single function
$\eta(vv')$.  Comparing the present case to the semileptonic one, we
shall use the same saturation assumption as in \cite{BM} and write
\begin{equation}\label{eqSat}
  \eta(vv') = |\xi(vv')|^2.
\end{equation}
For numerical calculations we use the measurement \cite{Barish95} of
the Isgur-Wise function
\begin{equation}
  \xi(vv') = 1 - a (vv' - 1), \qquad a = 0.84.
\end{equation}
Since the exclusive semileptonic decays are spectatorlike and since
in the limit of factorization, the matrix elements of the heavy quark
current are the same in the semi- and nonleptonic cases, dominance of
spectatorlike decays is a natural assumption for the nonleptonic
right charm case as well.  Therefore we use the results of
Sec.~\ref{secRC} for the channels
\begin{equation}
  \begin{array}{cccc}
    B^+ & \to & \overline D^0_{\rm dir} & X, \\
    B^+ & \to & \overline D^{*0} & X,
  \end{array}
  \qquad
  \begin{array}{cccc}
    B^0 & \to & D^-_{\rm dir} & X, \\
    B^0 & \to & D^{*-} & X,
  \end{array}
\end{equation} 
neglecting possible contributions from nonspectator channels such as
$B^+ \to D^- X$.  Another class of decays allow to obtain a
factorizable expression for $H_{\rm eff}$ after a Fierz
transformation.  However, the color indices need to be rearranged as
well, yielding a suppression by one power of $1/N_C$ in the amplitude.
Since we are working to leading order in $1/N_C$, we find a vanishing
rate for $B^0 \to \overline D^{*0} X$, whereas the channel $B^0 \to
\overline D^0 X$ is fed by the decay chain via the $D^{*-}$ meson.

For the contribution of the quark level decay $b \to c \overline c s$,
we assume the same number of right charm quarks to hadronize as
$\overline D^{(*)}$ as wrong charm quarks hadronize as either
$D^{(*)}$ or $D^{(*)}_s$.  Using the CLEO wrong charm measurement
${\rm Br}(B \to D X) = (7.9 \pm 2.2) \%$ \cite{Coan98} and the
Particle Data Group average ${\rm Br}(B \to D^\pm_s X) = (10.0 \pm
2.5) \%$ \cite{PDG} and neglecting a possible right charm $B \to
D^-_s$ contribution, we obtain a total right charm contribution from
$b \to c \overline c s$ of about $18 \%$.

It is known that the channel $b \to c \overline c s$ receives large
radiative corrections computed in \cite{Bagan95}.  We shall not
include these corrections in their detailed form, rather we shall take
into account their bulk effect by adjusting the charm quark mass in
Eqs.~(\ref{pheavy})--(\ref{rcfinal}).  Inserting an ``effective'' mass
$m_c^{\rm eff} = 1.0\,{\rm GeV}$ into the tree level relation, the
measured rate is reproduced.

Data are sparse for the wrong charm part, therefore we replace the
unknown functions in Eqs.~(\ref{factwc}) and (\ref{partwc}) by
\begin{deqnarray}
         F (v'Q,Q^2) & = \nc f \; 
                         2 \pi \delta \left( (Q-m_c v')^2 \right), \nl
  \tilde{F}(v'Q,Q^2) & = \nc \tilde{f} \;
                         2 \pi \delta \left( (Q-m_c v')^2 \right),
\end{deqnarray}%
where $f$ and $\tilde{f}$ are constants.  This parametrization using a
delta function is motivated by the negligible mass of the strange
quark against which the two charmed quarks recoil.  It can actually be
checked by measuring the invariant mass $m_{X'}$ of the additional
decay products in $B \to D \overline D X'$ which should turn out to be
small.  Setting $\tilde{f}$ to one, Eq.~(\ref{partwc}) reproduces the
well-known parton model result.  The fits yield
\begin{equation}
  f = 0.147 \, {\rm GeV} \qquad {\rm and} \qquad \tilde{f} = 0.121.
\end{equation}
Note that one naively expects $\tilde{f} = 1/8$ when assuming
isospin invariance in $B \to D^* X$ and $B \to D_{\rm dir} X$.  Under
this assumption, we have two charge states and four spin states, hence
in total eight states contributing.

For the semileptonic contributions, we use the results of
\cite{BM},\footnote{Note the sign error afflicting one term in
\cite{BM}, Eq.~(43), which should read\\
\( E_V^{B^0 \overline D^0}(y) =
   - \frac{C_{11}(y,\Lambda)}{C_3^2(y,\Lambda)}
     {\rm Br}(D^{*-} \to \overline D^0 X)
     \frac{1}{2} (y^2-1) |X(y)|^2
   + C_{18}(y,\Lambda) \frac{1}{N_C}.
\)}
but apply the same approximations as for the nonleptonic channels.
We neglect the renormalization group improvement and find
\begin{eqnarray}
  G_{sl}(M^2) & = &
    \frac{G_F^2}{2} |V_{cb}|^2 \; \frac{1}{3 \pi}
      (M^2 g_{\mu\nu} - M_\mu M_\nu) \Theta(M^2) \; \eta(vv') \\
  \nonumber &&
  {\rm Tr} [\overline H_B(v) \gamma^\mu (1-\gamma_5)
            H_{\overline D^{(*)}}(v')] \;
  {\rm Tr} [\overline H_{\overline D^{(*)}}(v')
            \gamma^\nu (1-\gamma_5) H_B(v)].
\end{eqnarray}
The $\tau$ contribution follows from
Eqs.~(\ref{pheavy})--(\ref{rcfinal}) replacing $V_{cs}$, $N_C$ and
$C_1$ by one and the charm quark mass by the $\tau$ mass.

Table~\ref{tabBR} shows the predictions for the total rates of these
two ans\"atze for the wrong charm piece.  For the numerical
calculation, we used $|V_{cb}|=0.04$.

\begin{table}
\renewcommand{\baselinestretch}{1}\normalsize
\begin{center}
\begin{tabular}{|l|r|r|r|l|}
\hline
Mode & Model 1 & Model 2 & \multicolumn{2}{c|}{Experiment} \\
\hline
$B \to \orbar{D} X$                          & $68.8\%$ & $68.8\%$ &
  $(87.2 \pm 3.5)\%$ & $\orbar{D}^0 + D^\pm$ \rule{0pt}{2.3ex} \\
$B \to \orbar{D}^* X$                        & $51.8\%$ & $52.7\%$ &
  $(48.7 \pm 3.1)\%$ & $\orbar{D}^{*0} + D^{*\pm}$ \\
$B \to \orbar{D}_{\rm dir} X$                & $17.1\%$ & $16.1\%$ &
  $(38.5 \pm 4.7)\%$ & $\orbar{D} - \orbar{D}^*$ \\
$B \to D X$                                  & $ 7.9\%$ & $ 7.9\%$ &
  $( 7.9 \pm 2.2)\%$ & \cite{Coan98} (input) \\
$B \to D^{*} X$                              & $ 5.0\%$ & $ 5.9\%$ && \\
\hline
$B \to \orbar{D}^0 X$                        & $52.0\%$ & $52.3\%$ &
  $(63.1 \pm 2.9)\%$ & \cite{PDG,Gibbons97B} \rule{0pt}{2.3ex} \\
$B \to D^{\pm}X$                             & $16.8\%$ & $16.5\%$ &
  $(24.1 \pm 1.9)\%$ & \cite{PDG,Gibbons97B} \\
$B \to \orbar{D}^{*0} X$                     & $25.9\%$ & $26.3\%$ &
  $(26.0 \pm 2.7)\%$ & \cite{PDG,Gibbons97B} \\
$B \to D^{*\pm} X$                           & $25.9\%$ & $26.3\%$ &
  $(22.7 \pm 1.6)\%$ & \cite{PDG,Gibbons97B,Albrecht96D} \\
$B \to \orbar{D}^0_{\rm dir} X$              & $ 8.5\%$ & $ 8.1\%$ &
  $(21.7 \pm 4.1)\%$ & $\orbar{D}^0 \!\!\! - \!\!
  \orbar{D}^{*0} \!\!\! - \!\! 0.68 D^{*\pm}$ \\
$B \to D^\pm_{\rm dir} X$                    & $ 8.5\%$ & $ 8.1\%$ &
  $(16.8 \pm 2.9)\%$ & $D^\pm - 0.32 D^{*\pm}$ \\
\hline
$B \to D^- \ell^+ \nu_\ell X$                & $ 2.0\%$ & $ 2.0\%$ &
  $( 2.7 \pm 0.8)\%$ & \cite{PDG,Fulton91} \\
$B \to \overline D^0 \ell^+ \nu_\ell X$      & $ 6.5\%$ & $ 6.5\%$ &
  $( 7.0 \pm 1.4)\%$ & \cite{PDG,Fulton91} \\
$B \to D^{*-} \ell^+ \nu_\ell X$             & $ 3.3\%$ & $ 3.3\%$ &
  $( 2.8 \pm 0.4)\%$ & \cite{PDG,Barish95} \\
$B \to \overline D^{*0} \ell^+ \nu_\ell X$   & $ 3.3\%$ & $ 3.3\%$ &
  $( 3.2 \pm 0.7)\%$ & \cite{PDG,Barish95} \\
$B \to D^{-} \tau^+ \nu_\tau X$              & $ 0.6\%$ & $ 0.6\%$ && \\
$B \to \overline D^{0} \tau^+ \nu_\tau X$    & $ 2.0\%$ & $ 2.0\%$ && \\
$B \to D^{*-} \tau^+ \nu_\tau X$             & $ 1.0\%$ & $ 1.0\%$ && \\
$B \to \overline D^{*0} \tau^+ \nu_\tau X$   & $ 1.0\%$ & $ 1.0\%$ && \\
\hline
$B^0 \to \overline D^0 X$                    & $31.8\%$ & $31.8\%$ && \\
$B^0 \to D^- X$                              & $29.1\%$ & $29.1\%$ && \\
$B^+ \to \overline D^0 X$                    & $60.9\%$ & $60.9\%$ && \\
$B^+ \to D^- X$                              & $ 0  \%$ & $ 0  \%$ && \\
$B^0 \to D^0 X$                              & $ 5.7\%$ & $ 6.0\%$ && \\
$B^0 \to D^+ X$                              & $ 2.2\%$ & $ 1.9\%$ && \\
$B^+ \to D^0 X$                              & $ 5.7\%$ & $ 6.0\%$ && \\
$B^+ \to D^+ X$                              & $ 2.2\%$ & $ 1.9\%$ && \\
$B^0 \to \overline D^{*0} X$                 & $ 0  \%$ & $ 0  \%$ && \\
$B^0 \to D^{*-} X$                           & $46.8\%$ & $46.8\%$ && \\
$B^+ \to \overline D^{*0} X$                 & $46.8\%$ & $46.8\%$ && \\
$B^+ \to D^{*-} X$                           & $ 0  \%$ & $ 0  \%$ && \\
$B^0 \to D^{*0} X$                           & $ 2.5\%$ & $ 3.0\%$ && \\
$B^0 \to D^{*+} X$                           & $ 2.5\%$ & $ 3.0\%$ && \\
$B^+ \to D^{*0} X$                           & $ 2.5\%$ & $ 3.0\%$ && \\
$B^+ \to D^{*+} X$                           & $ 2.5\%$ & $ 3.0\%$ && \\
\hline
\end{tabular}
\end{center}
\setlength{\abovecaptionskip}{7pt}
\caption{Comparison of our results with data.  Model~1
uses Eq.~(\ref{factwc}), model~2 uses Eq.~(\ref{partwc}).  Branching
ratios are computed using $\tau_{B^+} = \tau_{B^0} = 1.55 $ ps.}
\label{tabBR}
\end{table}

\section{Comparison with Data}
Data on one-particle inclusive $B$ decays are available from
$\Upsilon(4S)$ machines and also from the CERN electron positron
collider LEP. At the $\Upsilon(4S)$, the total production rates of
$D^{(*)+}$ and $D^{(*)-}$ or $\overline D^{(*)0}$ and $D^{(*)0}$ are
measured on the resonance, from which one can only deduce the rates
for $B$ admixture
{\renewcommand{\arraystretch}{2}
\begin{deqnarray}
  \Gamma (B \to D^{(*)\pm} X) & = \nc
    \frac{1}{2} \left[ \Gamma (B^+ \to D^{(*)\pm} X) +
    \Gamma (B^0 \to D^{(*)\pm} X) \right], \nl
  \Gamma (B \to \orbar{D}^{(*)0} X) & = \nc
    \frac{1}{2} \left[ \Gamma (B^+ \to \orbar{D}^{(*)0} X) +
    \Gamma (B^0 \to \orbar{D}^{(*)0} X) \right].
\end{deqnarray}}%
Table~\ref{tabBR} lists the available data. In the fourth row of the
table we list the CLEO measurement of wrong charm decays $B \to D X$
which is again the average over $B^+$ and $B^0$.

Although we expect our method to work best near the nonrecoil point
$vv'=1$, we integrate the spectra in order to obtain total rates,
assuming we can still get some insight into the bulk features.  Since
we calculate the right and wrong charm contributions separately, we
have to add them in order to obtain the number of $\orbar{D}$ mesons
produced in B meson decays.  This procedure corresponds to the usual
multiplicity definition of the branching ratio \cite[p.~570]{PDG}.

The first observable one can study is the total number of $\orbar{D}$
mesons in $B$ decays.  This quantity is considerably smaller than the
usually performed charm counting since it does not include the
production of $D_s$ mesons, charmed baryons and charmonium, the sum of
which amounts to about $20\%$ of the total charm production in $B$
decays.  Using the saturation assumption in Eq.~(\ref{eqSat}), our
model yields a total $\orbar{D}$ rate about $20\%$ below the
experimental result.  However, Eq.~(\ref{eqSat}) was taken from the
analysis of the semileptonic decays where the $\overline D$ and
$\overline D^*$ exclusive final states saturate only $(71 \pm 13)\%$
of the inclusive rate $B \to \overline D \ell^+ \nu_\ell X$.  Thus
Eq.~(\ref{eqSat}) should be replaced by
\begin{equation}
  \eta(vv') = |\xi(vv')|^2 + \delta \eta(vv'),
\end{equation}
where $\delta \eta$ accounts for the remaining contributions.  Because
we use factorization, this would enhance the nonleptonic channels by
a similar amount and hence improve the prediction for the total number
of $\overline D$ mesons.

Keeping this in mind, we still stick to the saturation assumption in
Eq.~(\ref{eqSat}) since $\delta \eta$ is unknown and the naive
use of Eq.~(\ref{eqSat}) reproduces the data on $B \to \orbar{D}^* X$
decays.  This could be accidental, as the real problem is spin
counting.  The ratio of vector to direct pseudoscalar mesons $r_S$
defined in Eq.~(\ref{spincount}) is expected to be about three, but
experiment yields a ratio barely above one (see Table~\ref{tabSC}).
Even in the semileptonic case, the experimental value is slightly
lower than expected, although errors are large.  Since the rates of
pseudoscalar and vector $\orbar{D}$ mesons are connected by heavy
quark spin symmetry, this large discrepancy is difficult to
understand.

\begin{table}
\begin{center}
\begin{tabular}{|l|r@{}c@{}l|r@{}c@{}l|}
\hline
$\displaystyle r_S =
  \frac{{\rm Br}(B \to \orbar{D}^* X) \rule{0pt}{2.3ex}}
       {{\rm Br}(B \to \orbar{D}_{\rm dir} X) \rule{0pt}{2.3ex}}$ &
  \multicolumn{3}{l|}{\begin{tabular}{l}All\\channels\end{tabular}} &
  \multicolumn{3}{l|}{\begin{tabular}{l}Semi-\\leptonic\end{tabular}} \\
\hline
Naive spin counting &
  \multicolumn{3}{l|}{3} & \multicolumn{3}{l|}{3} \\
Model 1 [see Eq.~(\ref{factwc})] & 3.04 & & & & & \\
Model 2 [see Eq.~(\ref{partwc})] & 3.26 & &
  & \raisebox{\height}[0pt]{3.36} & & \\
\hline
Fit all data using $\orbar{D}^{(*)0}$, $D^{(*)\pm}$ \rule{0pt}{2.3ex} &
  $1.39$&$\pm$&$0.27$ & $1.58$&$\pm$&$0.70$ \\
Data using $\orbar{D}^*$ and $\orbar{D}_{\rm dir}$ only &
  $1.26$ & $\pm$ & $0.22$ & $1.62$&$\pm$&$0.91$ \\
Data using $\orbar{D}^0$ and $D^\pm$ only &
  $1.92$ & $\pm$ & $0.46$ & $1.87$&$\pm$&$1.74$ \\
\hline
\end{tabular}
\end{center}
\caption{Predicted and measured spin counting ratio $r_S$.  For the
last line and for the fit, the relation~(\protect\ref{chargespin})
between charge and spin counting has been used as a constraint.  For
the column ``all channels,'' the error of the fit has been scaled by
$\sqrt{\chi^2/N_f}$, where $\chi^2 = 4.2$ and $N_f = 2$.}
\label{tabSC}
\end{table}

\begin{figure}
\begin{center}
\includegraphics{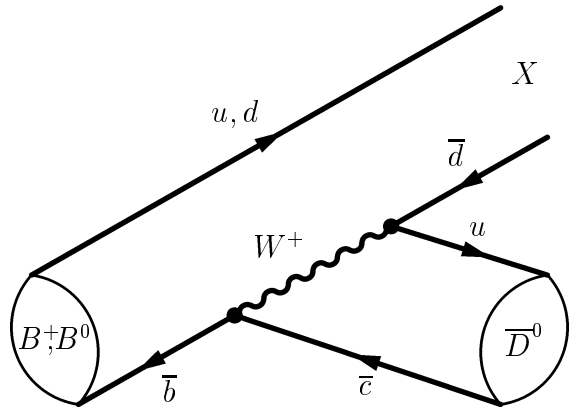}
\caption{Nonfactorizing isospin-violating topology, supplying a
possible explanation for an enhanced $\overline D^0$ rate.}
\label{figNF}
\end{center}
\end{figure}

Although the effect is less pronounced than the spin counting problem,
neutral $\orbar{D}$ mesons tend to occur slightly more often than
expected, see Table~\ref{tabQC}.  Assuming the measured value for the
spin counting ratio $r_S$, the measured charge counting ratios $r_Q$
are generally one or two standard deviations above expectations.  If
confirmed, this kind of effect would point to an additional
contribution which could for instance arise from a nonfactorizing
topology as depicted in Fig.~\ref{figNF}.  Nevertheless, current data
are still consistent with the relation~(\ref{chargespin}) between
charge and spin counting, see the fit in Table~\ref{tabSC}.

\begin{table}
\begin{center}
\begin{tabular}{|l|c|c|c|r@{}c@{}l|}
\hline
Channel \hfill Model & Naive & 1 & 2 &
  \multicolumn{3}{c|}{Data} \\
\hline \rule{0pt}{2.3ex}
${\rm Br}(B \to \orbar{D}^{*0} X) / {\rm Br}(B \to D^{*\pm} X)$ &
  $1$ & $1$ & $1$ & $1.15$&$\pm$&$0.14$ \\
${\rm Br}(B \to \orbar{D}^0_{\rm dir} X) /
  {\rm Br}(B \to D^\pm_{\rm dir} X)$ \rule{0pt}{2.3ex} &
  $1$ & $1$ & $1$ & $1.29$&$\pm$&$0.17$ \\
${\rm Br}(B \to \orbar{D}^0 X) / {\rm Br}(B \to D^\pm X)$ &
  $3$ & $3.09$ & $3.17$ & $2.62$&$\pm$&$0.24$ \\
\hline \rule{0pt}{2.3ex}
${\rm Br}(B \to \orbar{D}^{*0} \ell \nu X) /
  {\rm Br}(B \to D^{*\pm} \ell \nu X)$ &
  $1$ & \multicolumn{2}{c|}{$1$} & $1.14$&$\pm$&$0.30$ \\
${\rm Br}(B \to \orbar{D}^0_{\rm dir} \ell \nu X) /
  {\rm Br}(B \to D^\pm_{\rm dir} \ell \nu X)$ \rule{0pt}{2.3ex} &
  $1$ & \multicolumn{2}{c|}{$1$} & $1.05$&$\pm$&$0.89$ \\
${\rm Br}(B \to \orbar{D}^0 \ell \nu X) /
  {\rm Br}(B \to D^\pm \ell \nu X)$ &
  $3$ & \multicolumn{2}{c|}{$3.20$} & $2.59$&$\pm$&$0.93$ \\
\hline
\end{tabular}
\end{center}
\caption{Predicted and measured charge counting ratios $r_Q$.
Using the measured value $r_S \approx 3/2$, naive charge
counting yields $r_Q = 7/3$ instead of 3.}
\label{tabQC}
\end{table}

\begin{figure}
\begin{center}
\includegraphics{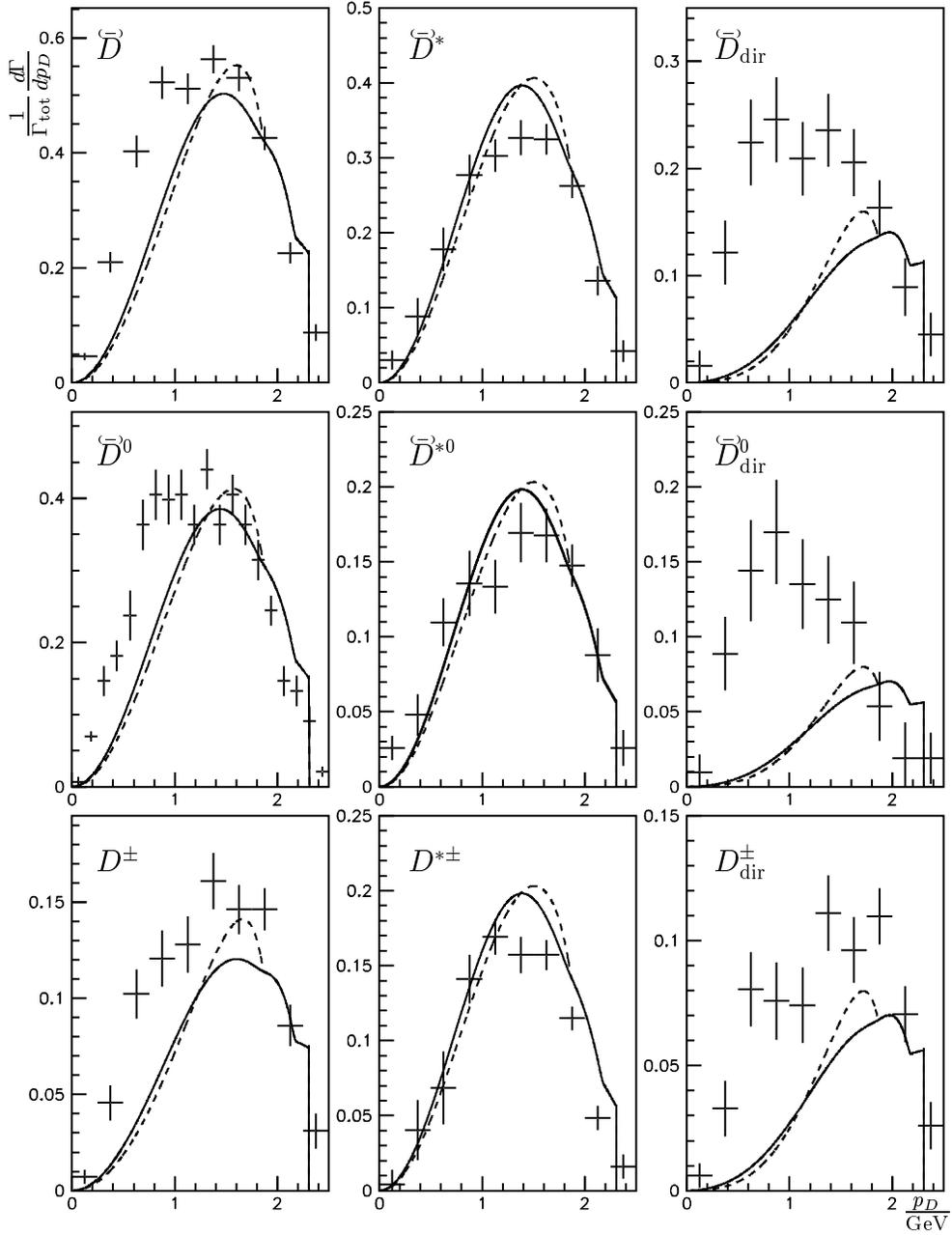}
\caption{Momentum spectra $1/\Gamma_{\rm tot} \: d\Gamma/dp_D$ of $B
\to \protect\orbar{D} X$ in comparison with theoretical predictions.
The solid line is model 2, the dashed one model 1.  The columns refer
to $\protect\orbar{D}$, $\protect\orbar{D}^*$ and direct
$\protect\orbar{D}$ mesons, the rows to a charge sum, neutral and
charged $\protect\orbar{D}^{(*)}$ mesons.}
\label{figSpect}
\end{center}
\end{figure}

Another piece of information for these decays are the $\orbar{D}$
momentum spectra.  These have been measured by ARGUS \cite{Albrecht91}
and CLEO \cite{Gibbons97B}, we use the data from CLEO which is more
recent and more precise.  The spectra are momentum distributions in
the rest frame of the $\Upsilon(4S)$.  However, the effect of the
motion of the $B$ mesons produces only a negligible smearing of these
spectra, and we can safely ignore this effect here.  In
Fig.~\ref{figSpect} we compare the data obtained by CLEO
\cite{Gibbons97B} with our theoretical prediction.

The spectra show that there is indeed a problem with the ratio of
pseudoscalar to vector mesons.  Using the saturation assumption in
Eq.~(\ref{eqSat}), the spectra of the vector mesons are described
within experimental uncertainties and the problem appears with the low
momentum region of the decay spectra for the pseudoscalar mesons.
Our theoretical ansatz, especially the SDE, should work best in this
region of small $\orbar{D}$ momentum.  In addition, the shape of the
spectra in this region is mainly determined by phase space, which
yields a behavior proportional to $p_D^2$ for small $p_D$ and constant
matrix element.  The steep rise of the momentum spectra for the
pseudoscalar $\orbar{D}$ mesons is thus difficult to understand.  An
investigation of slow direct pseudoscalar $\orbar{D}$ mesons,
i.e., those that do not originate from intermediate $\orbar{D}^*$
vector mesons, would be desirable.

\begin{figure}
\begin{center}
\includegraphics{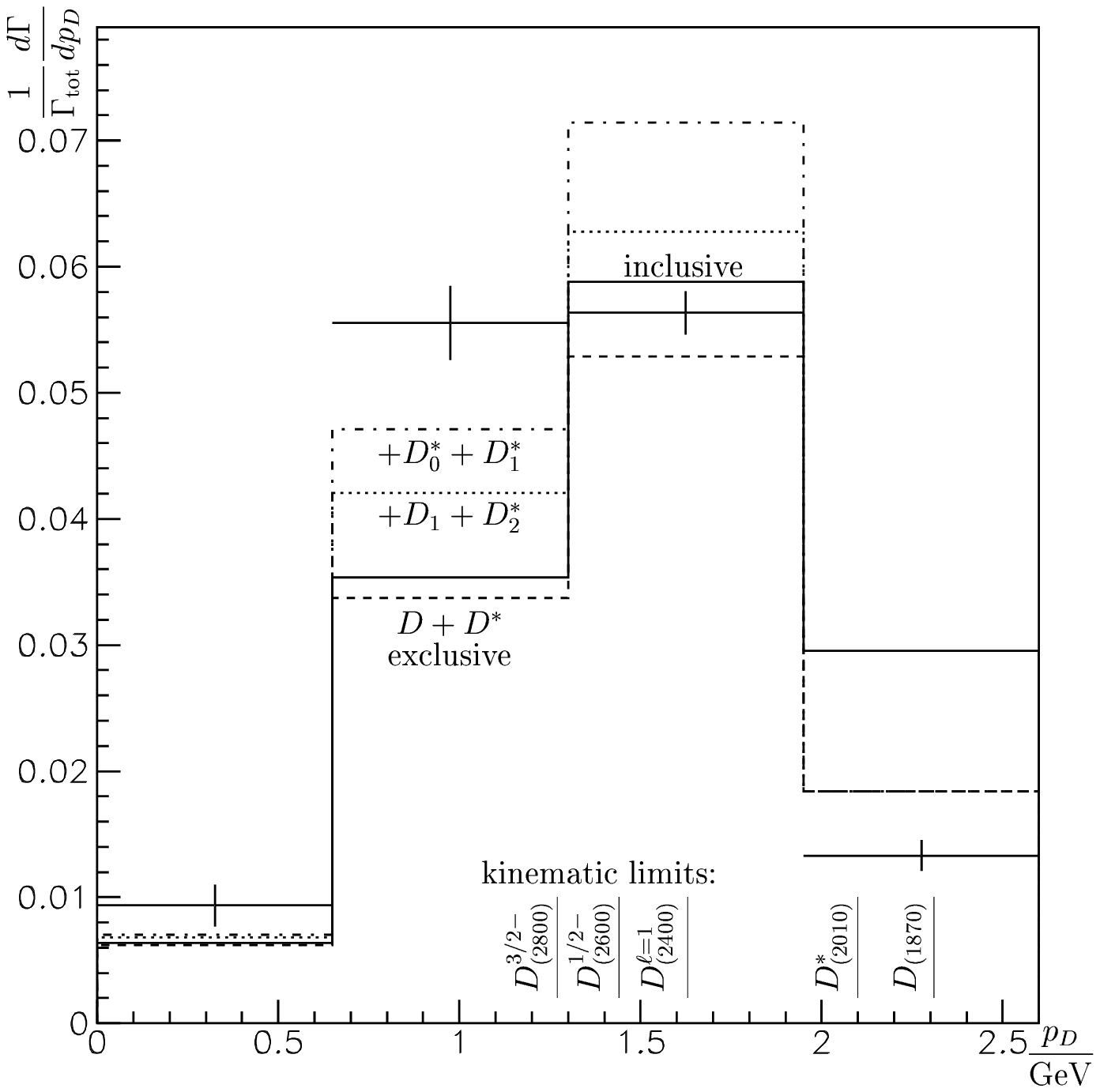}
\caption{One-particle inclusive semileptonic $\overline D$ momentum
spectrum in $B \to \overline D \ell^+ \nu_\ell X$ measured by CLEO
\protect\cite{Coan98}.  The solid line is the prediction of
\protect\cite{BM}; the dashed lines are sums of predictions for
exclusive channels following \protect\cite{Leibo98}.}
\label{figsl}
\end{center}
\end{figure}

This problem appears to be limited to nonleptonic decays.  In
Fig.~\ref{figsl} we compare the semileptonic $\overline D$ meson
momentum spectrum measured by CLEO \cite{Coan98} with the theoretical
predictions from \cite{BM} as well as with the sum of the first six
contributing exclusive channels \cite{Leibo98} and find no significant
disagreement.

\begin{figure}
\begin{center}
\includegraphics{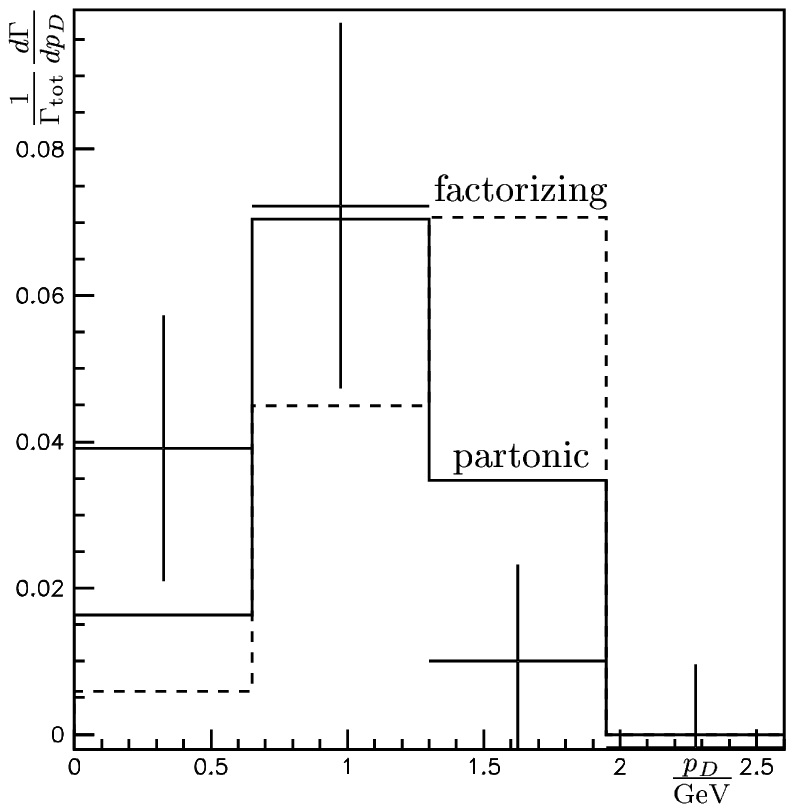}
\caption{One-particle inclusive wrong charm $D$ momentum spectrum in
$B \to D X$ measured by CLEO \protect\cite{Coan98} in comparison with
theoretical predictions.  The solid line is model 2; the dashed one
model 1.}
\label{figWCexp}
\end{center}
\end{figure}

There is also a measurement of the wrong charm $D$ spectrum by CLEO
\cite{Coan98}.  Unfortunately only four bins could be measured having
still substantial uncertainties.  In Fig.~\ref{figWCexp} we see that
the parton-model inspired model~2 fits the data better than model~1,
although evidence is not yet conclusive due to the quality of the
data.

\section{Conclusions}
In this paper we have developed a QCD based description of
one-particle inclusive decays of the type $B \to \overline D X$ and $B
\to D X$. The method we suggest is based on the large-$N_c$ limit of
QCD, allowing us to factorize certain matrix elements.  Once
factorization has been performed, one can identify pieces in the rates
which can be treated by a short distance expansion, assuming the
bottom and the charm quarks to be heavy.  This yields a series in
inverse powers of the parameter $m_b - m_c$.  The numerator of the
expansion parameter is a typical QCD scale for the light degrees of
freedom.  Thus the expansion parameters are $\Lambda_{QCD}/(m_b -
m_c)$, $1/N_C$ and $\alpha_s(m_c)$ and hence corrections to our
calculation could be fairly large, in the worst case of the order of
$30\%$.

We have studied the leading term of this expansion which still
contains a number of unknown nonperturbative functions. These
functions have to be parametrized.  In the same way as in the
semileptonic case, we reduce the number of functions appearing in the
right charm contributions to a single nonperturbative function which
can be related to the Isgur Wise function, once we assume that most of
the rate is saturated by the decays into the two ground state mesons
$\orbar{D}$ and $\orbar{D}^*$.  For the wrong charm case we suggest
two models corresponding to two different ways of contracting the
spinor indices.  Both models have a single nonperturbative form
factor, which we adjust to the experimental wrong charm yield.

Although our method should work best close to the nonrecoil point, we
also calculate total rates in order to discuss the total number of
$\orbar{D}$ mesons in $B$ decays and the spin and charge counting.
While the well-known problem of spin counting, see, e.g., the review
\cite{ARGUS}, is not solved by our ansatz, charge counting seems to
work well, once we properly take the $D^* \to D$ decays into account.

The shapes of the decay spectra into vector mesons are already
described quite satisfactorily, in particular in the low momentum
region, while the $\orbar{D}$ meson spectra are off in this region.
Furthermore, our model reproduces the normalization of the
$\orbar{D}^*$ spectra, such that the spin counting problem manifests
itself in a deficit of $\orbar{D}$ mesons.  Since the experimental
rates are above the theoretical ones, it would be interesting to
investigate which exclusive channels contribute in the small momentum
region.

\clearpage
\section*{Acknowledgments}
The authors thank U.~Nierste for discussions concerning the method,
I.~Bigi for valuable criticism and L.~Gibbons for clarifications
concerning the CLEO data.  We also thank C.~Balzereit who participated
in the early stages of this work and M.~Feindt for discussions on the
experimental prospects.  This work was supported by the DFG
Graduiertenkolleg ``Elementarteilchenphysik an Beschleunigern'' and by
the DFG Forschergruppe ``Quantenfeldtheorie, Computeralgebra und
Monte-Carlo-Simulation.''


\end{document}